\documentclass[12pt]{article}
\input{epsf}
\usepackage{epsf,cite,epsfig,psfig,float,amssymb,stmaryrd,latexsym}
\usepackage{graphics,psfrag}
\usepackage{a4wide,graphicx}
\usepackage{cite}

\newcommand{\nn}{\nonumber}
\newcommand{\be}{\begin{equation}}
\newcommand{\ee}{\end{equation}}
\newcommand{\ba}{\begin{eqnarray}}
\newcommand{\ea}{\end{eqnarray}}
\newcommand{\vs}{\vspace{-0.275cm}}

\begin{document}

\title{Chiral dynamics of the two $\Lambda(1405)$ states}

\author{
D. Jido$^{a,c}$, J.A. Oller$^b$, E. Oset$^{c}$, A. Ramos$^{d}$ and 
U.-G. Mei{\ss}ner$^e$ \\
{\small $^{a}$Research Center for Nuclear Physics (RCNP), Osaka University,}\\
{\small  Ibaraki, Osaka 567-0047, Japan}\\
{\small $^{b}$Departamento de F\'{\i}sica, Universidad de Murcia, 30071 Murcia, Spain}\\
{\small $^{c}$Departamento de F\'{\i}sica Te\'orica and IFIC,
Centro Mixto Universidad de Valencia-CSIC,} \\
{\small Institutos de
Investigaci\'on de Paterna, Aptd. 22085, 46071 Valencia, Spain}\\
{\small $^{d}$Departament d'Estructura i Constituents de la Mat\`eria,
Universitat de Barcelona,} \\
{\small Diagonal 647, 08028 Barcelona, Spain}\\
{\small $^{e}$ Universit\"at Bonn, Helmholtz-Institut f\"ur Strahlen- und
Kernphysik (Theorie)}  \\
{\small Nu{\ss}alle 14-16, D-53115 Bonn, Germany}\\
}

\date{\today}

\maketitle
\begin{abstract}
 Using a chiral unitary approach for the meson--baryon interactions, we show
 that two octets of $J^{\pi}=1/2^-$  baryon states, which are
degenerate in the limit of
 exact SU(3) symmetry, and a singlet are generated dynamically.  The SU(3)
 breaking produces the splitting of the two octets,  resulting in the case of
strangeness $S=-1$ in two poles of the scattering matrix close to the nominal
$\Lambda(1405)$ resonance. These poles are combinations of the singlet state
 and the octets.  We show how actual experiments  see just one effective
resonance shape, but with properties which change from one reaction to
 another.
\end{abstract}

\newpage

\section{Introduction}
The $\Lambda(1405)$ resonance has been a long-standing example of a dynamically
generated resonance appearing naturally in scattering theory with coupled 
meson--baryon channels with strangeness $S=-1$ \cite{dalitz}. Modern chiral formulations
of the meson--baryon interaction within unitary frameworks all lead to the
generation of this resonance, which is seen as a near Breit--Wigner form in the
 mass distribution of $\pi \Sigma$ states with isospin $I=0$ in hadronic production processes
 \cite{norbert,angels,oller,phase}. Yet, it was shown that in some models one
could obtain two poles close to the nominal $\Lambda(1405)$
resonance, as it was the case within the cloudy bag model in
Ref.~\cite{fink}. Also, in the investigation of the poles of the
 scattering matrix in Ref.~\cite{oller}, within the
 context of chiral dynamics, it was found that
 there were two poles close to the nominal $\Lambda(1405)$ resonance both
 contributing to the $\pi\Sigma$ invariant mass distribution.  This
 was also the case in Refs.~\cite{magnetic,nieves}, where
 two poles are obtained with similar properties as to their masses, widths and
 partial decay widths compared to those of the previous works. 

 The purpose of this paper is to investigate further  the 
 origin
 of these two poles, their nature and why there seems to be only one resonance
 in actual experiments. We furthermore suggest  new experiments which could reveal the
 presence of these two states.

The manuscript is organized as follows. In Section~\ref{sec:2}, we briefly summarize the
salient features of the chiral unitary coupled channel approach for the interactions between
the octet of Goldstone bosons and the octet of the lowest baryons. The poles in the corresponding
meson--baryon scattering matrix are discussed in Section~\ref{sec:3}, with particular emphasis on 
the group structure of the dynamically generated resonances in the SU(3) limit. The relation between
these states and their couplings in that limit and in the physical case is further elaborated on in
 Section~\ref{sec:4}. In Section~\ref{sec:5}, we show how these poles manifest themselves in 
physical observables, and how different experiments are able to unravel the two poles generating the
$\Lambda (1405)$. Some conclusions and an outlook are given in Section~\ref{sec:6}. Some technicalities
are relegated to the appendices.

\section{Description of the meson baryon interactions}
\label{sec:2}
    Starting from the chiral Lagrangians for meson--baryon interactions
\cite{chiral} and using the N/D method to obtain a scattering matrix fulfilling
exactly unitarity in coupled channels \cite{oller}, the full set of transition matrix
elements with the coupled channels in  $S=-1$, $K^- p$,
 $\bar{K}^0 n$, $\pi^0 \Lambda$, $\pi^0 \Sigma^0$,
$\pi^+ \Sigma^-$, $\pi^- \Sigma^+$, $\eta \Lambda$, $\eta
\Sigma^0$, $K^0\Xi^0$ and
$K^+\Xi^-$,  is given in matrix form by

\begin{equation}
T = [1 - V \, G]^{-1}\, V~.
\label{eq:bs1}
\end{equation}
Here, the matrix $V$, obtained from the lowest order meson--baryon
chiral Lagrangian, contains the Weinberg-Tomozawa or seagull
contribution, as employed e.g.  in Ref.~\cite{bennhold},

\begin{equation}
%       V_{i j} = - C_{i j} \frac{1}{4 f^2} (k^0 + k'^0) \, .
V_{i j} = - C_{i j} \frac{1}{4 f^2}(2\sqrt{s} - M_{i}-M_{j})
\left(\frac{M_{i}+E}{2M_{i}}\right)^{1/2} \left(\frac{M_{j}+E^{\prime}}{2M_{j}}
\right)^{1/2}~ ,
\label{eq:ampl2}
\end{equation}
where the $C_{i j}$ coefficients are given in Ref.~\cite{angels},
and an averaged meson decay constant $f=1.123f_{\pi}$ is used
\cite{bennhold}, with $f_\pi = 92.4\,$MeV the weak pion decay constant.
 At lowest order in the chiral expansion all the
baryon masses are equal to the one in the chiral limit, $M_0$,
nevertheless in Ref.~\cite{bennhold} the physical baryon masses,
$M_i$, were used and these are the ones appearing in
Eq.~(\ref{eq:ampl2}). In addition to the Weinberg-Tomozawa term,
one also has at the same order in the chiral expansion the direct
and exchange diagrams considered in Ref.~\cite{oller}. These are
suppressed at low energies by powers of the three-momenta and
meson masses over $M_{i}$, the leading one being just linear.
However, their importance increases with energy and around
$\sqrt{s}\simeq 1.5$ GeV they can be as large as a $20\%$ of the
seagull term.

The diagonal
matrix $G$ stands
for the loop function of a meson and a baryon and is defined by a dispersion
relation in terms of phase space with a cut starting at the corresponding
threshold $s_l$, namely \cite{oller}:
\begin{eqnarray}
G(s)_l &=&G(s_0)_l-\frac{s-s_0}{\pi}\int_{s_l}^\infty ds'
\frac{\rho(s')_l}{(s'-s-i0^+)(s'-s_0)}~,\\
\rho(s)_l &=&\frac{M_i q_l}{4\pi\sqrt{s}}\label{defql} 
\end{eqnarray}
and
$G(s_0)_l$ is a subtraction constant. The above expression
corresponds to the loop function of a meson and a baryon once the
logarithmic divergent constant is removed:
\begin{eqnarray}
G_{l} &=& i \, \int \frac{d^4 q}{(2 \pi)^4} \, \frac{M_l}{E_l
(\vec{q}\,)} \,
\frac{1}{k^0 + p^0 - q^0 - E_l (\vec{q}\,) + i \epsilon} \,
\frac{1}{q^2 - m^2_l + i \epsilon} ~.
\label{gloop}
\end{eqnarray}
The analytical properties of $G$ are properly kept when evaluating
the previous loop function in dimensional regularization.
Using dimensional regularization and removing the divergent constant piece leads to
\begin{eqnarray}
G_{l} &=& i \, 2 M_l \int \frac{d^4 q}{(2 \pi)^4} \,
\frac{1}{(P-q)^2 - M_l^2 + i \epsilon} \, \frac{1}{q^2 - m^2_l + i
\epsilon}  \nonumber \\ &=& \frac{2 M_l}{16 \pi^2} \left\{ a_l(\mu) + \ln
\frac{M_l^2}{\mu^2} + \frac{m_l^2-M_l^2 + s}{2s} \ln \frac{m_l^2}{M_l^2} +
\right. \nonumber \\ & &  \phantom{\frac{2 M}{16 \pi^2}} +
\frac{q_l}{\sqrt{s}}
\left[
\ln(s-(M_l^2-m_l^2)+2 q_l\sqrt{s})+
\ln(s+(M_l^2-m_l^2)+2 q_l\sqrt{s}) \right. \nonumber  \\
& & \left. \phantom{\frac{2 M}{16 \pi^2} +
\frac{q_l}{\sqrt{s}}}
\left. \hspace*{-0.3cm}- \ln(-s+(M_l^2-m_l^2)+2 q_l\sqrt{s})-
\ln(-s-(M_l^2-m_l^2)+2 q_l\sqrt{s}) \right]
\right\}~ ,
\label{eq:gpropdr}
\end{eqnarray}
where $\mu$ is the scale of dimensional regularization. For a given value of this scale, the
subtraction constant $a_i (\mu) $ is determined so that the results are
finally scale independent.

The loop function represented by Eq.~(\ref{gloop}) was calculated
in Ref.~\cite{angels} with a cut-off regularization, similarly as
previously done in meson--meson scattering \cite{npa}.
%
%\begin{eqnarray}
%G_{l}&=& \int^{q_{\rm max}} \, \frac{d^3 q}{(2 \pi)^3} \, \frac{1}{2
%\omega_l(\vec{q}\,)}
%\,
%\frac{M_l}{E_l (\vec{q}\,)} \,
%\frac{1}{p^0 + k^0 - \omega_l (\vec{q}\,) - E_l (\vec{q}\,) + i \epsilon}~,
%\label{eq:gprop}
%\end{eqnarray}
%where $q_{max}$ is the three-momentum cut off.
The values of the $a_{i}$
 constants in Eq.~(\ref{eq:gpropdr})
are found to be around $-2$ to agree with the results of
the cut--off method for cut--off values of the order of the mass of
the $\rho(770)$ \cite{oller}, which we call of natural size.
Indeed, in Ref.~\cite{bennhold} it was found that with the values
for the subtraction constants
\begin{equation}
\begin{array}{lll} a_{{\bar K}N}=-1.84~~ &
a_{\pi\Sigma}=-2.00~~ & a_{\pi\Lambda}\,=-1.83 \\ a_{\eta
\Lambda}\,\,=-2.25~~ & a_{\eta\Sigma}=-2.38~~ & a_{K\Xi}=-2.67 \ .
\end{array}
\label{eq:coef}
\end{equation}
one reproduces the results for the $G$ functions obtained in
Ref.~\cite{angels} with a cut--off of 630 MeV.

It is further interesting to consider the large $N_c$ counting of the different terms 
present in  Eq.(\ref{eq:bs1}). From Eq.~(\ref{eq:ampl2}), since 
$f^2\propto N_c$ and also $M_{B_i}\propto N_c$, it follows  
that $V_{ij}$ is ${\cal O}(N_c^{-1})$ in
the large $N_c$ counting. 
Here, one has to expand
Eq.~(\ref{eq:ampl2}) in the limit of $M_{i}\rightarrow \infty$, 
taking into account
that $\sqrt{s}=\sqrt{q^2+M_{i}^2}+\sqrt{q^2+m_i^2}$, so that the
baryon
masses disappear when subtracted from  $2\sqrt{s}$ in the first term
between brackets of this equation which is then ${\cal O}(N_c^0)$. 
Since the other terms between brackets are also  ${\cal O}(N_c^0)$ 
it follows that $V_{ij}$ is ${\cal O}(N_c^{-1})$ because of the factor of 
$f^2$ appearing in the denominator. In the same way, by expanding 
Eq.~(\ref{eq:gpropdr})
in the limit $M_l\rightarrow \infty$, a term linear in $M_l$ appears
so that $G_l$ seems to be ${\cal O}(N_c)$. This term is energy-independent and is 
 given by the combination:
\be
\label{largencleading}
\frac{M_l}{8\pi^2}\left(a_l(\mu)+\ln\frac{M_l^2}{\mu^2}\right)~.
\ee
The energy-dependent terms are ${\cal O}(N_c^0)$. In ref.\cite{oller} it was 
shown that
\be
a_l(\mu)=-2 \ln\left(1+\sqrt{1+\frac{M_l^2}{\mu^2}}\right)+{\cal O}(\frac{1}{M_l})~,
\label{relation}
\ee
where the scale $\mu$ can be chosen such that it corresponds
to a hypothetical three-momentum cut-off with a natural value 
around $M_\rho$ used 
to evaluate $G_l(s)$, as e.g. in ref.\cite{angels}. Taking the limit $M_l\rightarrow 
\infty$ in this equation one then has:
\be
 a_l(\mu)\rightarrow- \ln \frac{M_l^2}{\mu^2}~,
\ee
and hence the leading combination of Eq.(\ref{largencleading}) reduces to an 
${\cal O}(N_c^0)$ contribution as the rest of the terms so that in this case $G_l$ is finally 
${\cal O}(N_c^0)$. This is also
the  order one would infer naturally from Eq.(\ref{defql}) by simply applying the 
scaling properties  $M_l\backsim {\cal O}(N_c)$ and $q \backsim {\cal O}(N_c^0)$ to 
the integral 
and then taking the accompanying subtraction constant $G_l(s_0)$ of the same order as that  
of the integral. Since from the studies of refs.\cite{angels,oller,bennhold} it is clear 
that one can consider the subtractions constant $a_l$ as originating from a cut-off as 
in Eq.(\ref{relation}), then we infer that the $G_l(s)$ function must be counted as 
${\cal O}(N_c^0)$. The important
point for us is that then the product $V \, G$, appearing in Eq.~(\ref{eq:bs1}), 
is ${\cal O}(N_c^{-1})$ and then suppressed in the large $N_c$ limit with respect to the 
identity. This situation is similar to that of the meson--meson case, 
where $V\, G$ is as well ${\cal O}(N_c^{-1})$\cite{nsd}. Thus, the dynamically generated
resonances disappear in the limit of large number of colours. Were the subtraction 
constants  not 
generated through a relation like that of Eq.(\ref{relation}), which implies a value of 
$a_l$ around $-2$, then $V\, G$ would be ${\cal O}(N_c^0)$ and would not be suppressed as 
compared to the identity so that the poles would survive the 
large $N_c$ limit. Thus, as a result of this discussion, we want to emphasize that 
the set of  resonances with $S=-1$ generated dynamically within our approach and 
to be presented in detail below, are suppressed in the large $N_c$ limit and are not 
suited to a large $N_c$ expansion as that employed recently in ref.\cite{goity}.

%Since along this work we are going to discuss  mainly
%refs.\cite{oller,bennhold}, we express the relations between their normalizations:
%\begin{eqnarray}
%T'_{ij}&=&-\sqrt{4M_{B_i}M_{B_j}}T_{ij}~,\nonumber\\
%V'_{ij}&=&-\sqrt{4M_{B_i}M_{B_j}}V_{ij}~,\nonumber\\
%G'_l&=&G_l/2M_{B_l}~,
%\end{eqnarray}
%where the quantities with primes correspond to Ref.~\cite{oller} and those
%without primes to Ref.~\cite{bennhold}.

\section{Poles of the T-matrix}
\label{sec:3}

  The study of Ref.~\cite{bennhold} showed the presence of  poles in
  Eq.~(\ref{eq:bs1}) around the $\Lambda(1405)$ and the
$\Lambda(1670)$ for isospin $I=0$ and around the $\Sigma(1620)$ in
$I=1$.  The same approach in $S=-2$ leads to the resonance
$\Xi(1620)$ \cite{xi} and in $S=0$ to the $N^*(1535)$
\cite{inoue}, this latter one also generated dynamically in
Ref.~\cite{siegel}.
 One is thus tempted to consider the appearance of a singlet and an octet
of meson--baryon resonances. Nevertheless, the situation is more
complicated because indeed in the SU(3) limit there are {\it two} octets
and not just one, as we discuss below. As a matter of fact, the
$\Lambda(1405)$ is a mixture of a singlet and an octet, and not just a
singlet as assumed in Ref.~\cite{goity}.
The presence of these multiplets was already
  discussed in Ref.~\cite{oller} after obtaining poles with $S=-1$ in the
$I=1$ channel,
with mass around 1430 MeV, and two poles with $I=0$, of masses around that of the
  $\Lambda(1405)$.
%Two of the three poles become degenerate in the $SU(3)$
%limit (octet members) and still remains a singlet pole..
  Similar ideas have been exploited in the meson--meson interaction where a
nonet of dynamically generated mesons, made of  the $\sigma(500)$, $f_0(980)$,
$a_0(980)$ and $\kappa(900)$, has been obtained \cite{npa,ramonet,nsd,norbert2}.

 The appearance of a multiplet of dynamically generated mesons and baryons seems
 most natural once a state of the multiplet appears. Indeed, one must recall that the
 chiral Lagrangians are obtained from the combination of the octet of
 pseudoscalar mesons (the pions and partners) and the octet of stable baryons
(the nucleons and partners).  The SU(3) decomposition of the combination of two
 octets tells us that
 \begin{equation}
 8 \otimes 8=1\oplus 8_s \oplus 8_a \oplus 10 \oplus \overline{10} \oplus 27~.
\end{equation}
Thus, on pure SU(3) grounds, should we have a SU(3) symmetric Lagrangian,
 one can expect e.g. one singlet and two octets of resonances, the symmetric and
 antisymmetric ones.  Actually in the case of the meson--meson interactions only
 the symmetric octet appears in S-wave because of Bose statistics, 
but in the case of the meson--baryon
 interactions, where the building blocks come from two octets of different
 nature, both the symmetric and antisymmetric octets could appear and there is
 no reason why they should be degenerate in principle.

 The lowest order of the meson--baryon chiral Lagrangian is exactly SU(3) 
invariant if
 all the masses of the mesons, or equivalently the quark
 masses,  are set equal. As stated above  [see Eq.~(\ref{eq:ampl2})], 
in Ref.~\cite{bennhold}
 the baryon masses take their physical values, although strictly
 speaking at the leading order in the chiral expansion they should be equal to
 $M_0$. For Eq.~(\ref{eq:ampl2}) being SU(3) symmetric, all the baryons masses
 $M_{i}$ must be set equal as well. When all the meson and baryon masses are
  equal, and these common masses are employed in evaluating the $G_l$ functions,
 together with equal subtraction constants $a_l$, the $T$--matrix obtained
from Eq.~(\ref{eq:bs1}) is also SU(3) symmetric. In appendix A we
show that in the SU(3) limit the subtraction constants $a_l$ are
independent of the
 physical channel.

If we do such an SU(3) symmetry approximation
and look for poles of the scattering matrix, we find poles
corresponding to the octets and singlet. The surprising result is that
the two octet poles are degenerate as a consequence of the 
 dynamics contained in
 the chiral Lagrangians. Indeed, if we evaluate the matrix elements of the transition potential
$V$ in a basis of SU(3) states,

\begin{equation}
V_{\alpha  \beta}=\sum_{i,j} \langle i, \alpha \rangle C_{ij}
\langle j , \beta \rangle,
\end{equation}
where $\langle i, \alpha\rangle$ are the SU(3) Clebsch--Gordan
coefficients and $C_{ij}$ the coefficients in
Eq.~(\ref{eq:ampl2}), we obtain:

\begin{equation}
V_{\alpha  \beta}= {\rm diag}(6,3,3,0,0,-2)~, \label{eq:su3}
\end{equation}
taking the following order for the irreducible representations:
$1$, $8_s$, $8_a$, $10$, $\overline{10}$ and $27$.
% YYY {\it In SU(3) the
%  isospin is like the third component of isospin in $SU(2)$ and the
%  irreducible matrix elements of the Wigner Eckart theorem do not depend on
%  it. They depend only of the SU(3) representation.}

Hence we observe that the states belonging to different
irreducible representations do not mix, and the scattering amplitude
of the two
octets is the same at Born level (i.e. to leading order in the
chiral expansion). Thus the two octets appear
degenerate after the unitarized resummation. The coefficients in
Eq.~(\ref{eq:su3}) clearly illustrate why there are no bound
states in the $10$, $\overline{10}$ and $27$ representations. 
Indeed, considering the minus sign in Eq.~(\ref{eq:ampl2}), a negative
sign in Eq.~(\ref{eq:su3}) means repulsion.
In practice, the same chiral Lagrangians allow for SU(3) breaking. In
the case of Refs.~\cite{angels,bennhold} the breaking appears
because both in the $V_{i j}$ transition potentials as in the $G_l$
loop functions one uses the
physical masses of the particles as well as different subtraction constants in $G_l$, 
corresponding to the use of a unique cut-off in all channels. 
 In Ref.~\cite{oller} the
physical masses are also used in the $G_l$ functions, although these functions are evaluated 
with a unique subtraction constant as corresponds to the SU(3) limit, see appendix A. In addition 
 the $V_{ij}$ transition potentials are evaluated strictly at lowest order in the chiral expansion 
so that a common baryon mass is used and the one baryon exchange diagrams, both direct and crossed, 
are included. In both approaches, physical masses are
used to evaluate the $G_l$ loop functions so that unitarity is fulfilled
exactly and the physical thresholds for all channels are respected. This is
important in a multichannel problem like the present case. 
In general terms, additional SU(3) breaking contributions can be included systematically  
when evaluating the interacting kernel $V$ in a chiral expansion of the meson-baryon interactions, 
as explained in ref.\cite{oller}. The lowest order SU(3) breaking corrections occur when $V$ 
is evaluated at next-to-leading order from the ${\cal O}(p^2)$ meson-baryon Lagrangian, 
with some of these terms already included in Ref.\cite{hosaka}. Other SU(3) breaking effects 
as e.g. those arising from the difference between the weak decay constants of the pseudoscalars 
appear at ${\cal O}(p^3)$ in $V$. The systematic inclusion of such higher order corrections is 
beyond the present study but should be considered in the future.

\begin{figure}
  \epsfxsize=14cm
  \centerline{\epsfbox{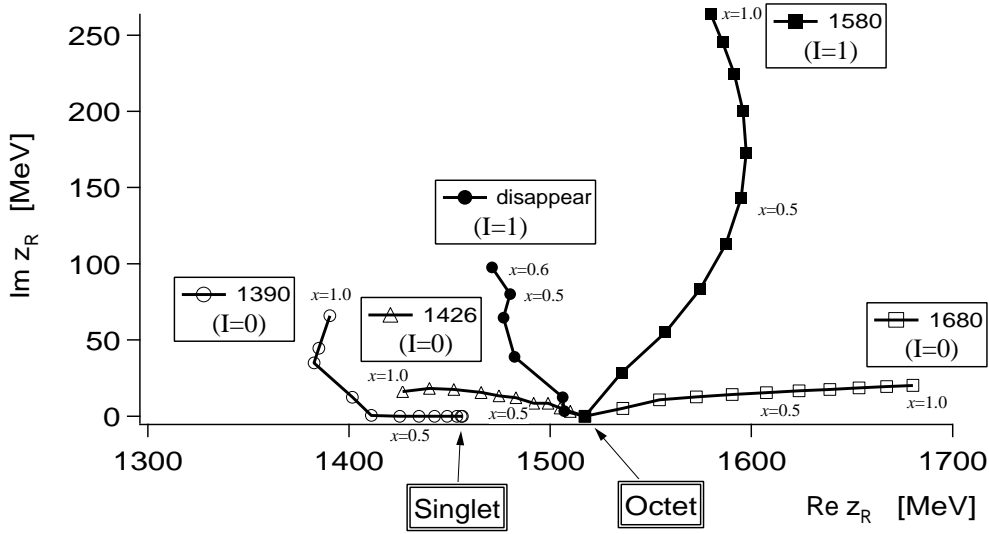}}
  \caption{Trajectories of the poles in the scattering amplitudes obtained by
  changing the SU(3) breaking parameter $x$ gradually. At the SU(3) symmetric 
  limit ($x=0$),
   only two poles appear, one is for the singlet and the other for the octet.
  The symbols correspond to the step size $\delta x =0.1$.
  \label{fig:tracepole}}
\end{figure}

By following the approach of Ref.~\cite{bennhold} and using the
physical masses of the baryons and the mesons, the position of the
poles change and the two octets split apart in four branches, two
for $I=0$ and two for $I=1$, as one can see in
Fig.~\ref{fig:tracepole}. In the figure we show the trajectories
of the poles as a function of a parameter $x$ that breaks
gradually the SU(3) symmetry up to the physical values.  The
dependence of masses and subtraction constants on the parameter
$x$ is given by
\begin{eqnarray}
M_i(x) &= & M_0+x(M_i-M_0),  \nonumber \\
m^{2}_{i}(x) &=& m_{0}^{2} + x (m^{2}_{i}-m^{2}_{0}), \nn\\
a_{i}(x) &=& a_{0} + x (a_{i} - a_{0}),
\end{eqnarray}
where $0\le x \le 1$. For the baryon masses, $M_{i}(x)$, the
breaking of the SU(3) symmetry follows linearly, while for the
meson masses, $m_{i}(x)$, the law is quadratic in the masses,
since in the QCD Lagrangian the flavor SU(3) breaking appears in
the quark mass terms and the squares of the meson masses depend on
the quark masses linearly.  In the calculation of
Fig.~\ref{fig:tracepole}, the values $M_{0}=1151$ MeV, $m_{0} =
368$ MeV and $a_{0}= -2.148 $ are used.

  The complex poles, $z_R$, appear in unphysical sheets. In the
present search we follow the strategy of changing
the sign of the momentum $q_l$ 
in the $G_l(z)$ loop function of
  Eq.~(\ref{eq:gpropdr}) for the channels which
are open at an energy equal to Re($z$).
%YYY {\em This is not right. One only changes from
%a Riemann sheet to another when crossing a cut, which is not the
%case simply when Re($z$) is larger than the threshold energy.}

The splitting of the two $I=0$ octet states is very interesting.
One moves to higher energies to merge with the $\Lambda(1670)$
resonance and the other one moves to lower energies to create a
pole, quite well identified below the  $\bar{K}N$ threshold, with
a narrow width. In appendix B we show explicitly 
how for small values of the symmetry breaking parameter $x$
the poles move away from their values in  the SU(3) symmetry limit.
We should also note that when for some values of
$x$ the trajectory crosses the $\bar{K}N$ threshold ($\sim 1435$
MeV) the pole fades away but it emerges again clearly for values
of $x$ close to $1$. On the other hand, the singlet also evolves
to produce a pole at low energies with a quite large width.

We note that the singlet and the $I=0$ octet states appear
nearby in energy and one of the purposes of this paper is,
precisely, to point out the fact that what experiments  actually
see is a combination of the effect of these two resonances.
This should be considered the main result of this paper.

Similarly as for the $I=0$ octet states, we can see that one branch of the 
$I=1$ states moves to higher energies while another
moves to lower energies. The branch moving to higher energies finishes at
what would correspond to the $\Sigma(1620)$ resonance when the physical
masses are reached. The
branch moving to lower energies fades away after a while when getting close to
the $\bar{K}N$ threshold.  

The model of Ref.~\cite{oller} reproduces qualitatively the same results.
However, this model also produces 
in the physical limit ($x=1$) another $I=1$ pole having Re($z$)=1401~MeV if,
in addition to changing the signs of
the on-shell momenta in the $\pi\Lambda$ and $\pi\Sigma$ channels
in accordance to the strategy mentioned above, the sign in the
${\bar K} N$ channel is also changed. This is legitimate in this case due to 
the proximity of the $\bar{K}N$ threshold to the position of the $I=1$ 
resonance of ref.\cite{oller}  as compared to its 
width (see the later table \ref{tab:oller0}). As we can see, the appearance of this pole
is more sensitive to the details of the coupled channel approach and hence it is not 
so stable as the $I=0$ poles discussed above.
Yet, the pole found in the model of Ref.~\cite{bennhold} 
for values of $x$ below to 0.6, see Fig.~\ref{fig:tracepole}, 
reflects, in the physical limit $x=1$, into a
strong cusp structure of the $I=1$ amplitudes at 
the $\bar{K}N$ threshold, leading to a pronounced
peak on top of a large background. In the model of
Ref.~\cite{oller}, the presence of a pole
in the Riemann sheet where the sign of the momenta
in the three channels ($\pi\Lambda$, $\pi\Sigma$, ${\bar K}N$) is changed
enhances the structure of this peak, which shows the features of
a resonant shape. Whether this enhancement in the $I=1$ amplitudes
can be interpreted as a 
resonance or as a cusp, 
the fact that the
strength of the $I=1$ amplitude around the
$\Lambda(1405)$ region is not negligible should have consequences for 
reactions producing $\pi \Sigma$ pairs in that region.
This has been illustrated for instance
in Ref.~\cite{nacher1}, where the photoproduction of the $\Lambda(1405)$ via the
reaction $\gamma p \to K^+ \Lambda(1405)$ was studied. It was shown there that
the different sign in the
$I=1$ component of the $\mid \pi^+ \Sigma^-\rangle$, $\mid \pi^- \Sigma^+\rangle$
states leads, through interference between the $I=1$ and the dominant $I=0$
amplitudes, to different
cross sections in the various charge channels, a fact that has been
confirmed experimentally very recently \cite{ahn}.

Once the pole positions are found, one can also determine the
couplings of these resonances to the physical states by studying
the amplitudes close to the pole and identifying them with
\begin{equation}
T_{i j}=\frac{g_i g_j}{z-z_R}~.
\end{equation}
The couplings $g_i$ are in general complex valued numbers.
In Tables~\ref{tab:jido0} and \ref{tab:jido1} we summarize the
pole positions and the complex couplings $g_i$ obtained from the
model of Ref.~\cite{bennhold} for isospin $I=0$ and $I=1$,
respectively.

\begin{table}[ht]
\centering \caption{\small Pole positions and couplings to $I=0$
physical states from the model of Ref.~\cite{bennhold}}
 \vspace{0.5cm}
\begin{tabular}{|c|cc|cc|cc|}
\hline
 $z_{R}$ & \multicolumn{2}{c|}{$1390 + 66i$} &
\multicolumn{2}{c|}{$1426 + 16i$} &
 \multicolumn{2}{c|}{$1680 + 20i$}  \\
 $(I=0)$ & $g_i$ & $|g_i|$ & $g_i$ & $|g_i|$ & $g_i$ & $|g_i|$ \\
 \hline
 $\pi \Sigma$ & $-2.5-1.5i$ & 2.9 & $0.42-1.4i$ & 1.5 & $-0.003-0.27i$ &
 0.27 \\
 ${\bar K} N$ & $1.2+1.7i$ & 2.1 & $-2.5+0.94i$ & 2.7 & $0.30+0.71i$ &
 0.77 \\
 $\eta\Lambda$ & $0.010+0.77i$ & 0.77 & $-1.4+0.21i$ & 1.4 & $-1.1-0.12i$ &
 1.1 \\
 $K\Xi$ & $-0.45-0.41i$ & 0.61 & $0.11-0.33i$ & 0.35 & $3.4+0.14i$ &
 3.5 \\
 \hline
 \end{tabular}
\label{tab:jido0}
\end{table}

\begin{table}[ht]
\centering \caption{\small Pole position and couplings to $I=1$
physical states from the model of Ref.~\cite{bennhold}}
\vspace{0.5cm}
\begin{tabular}{|c|cc|}
 \hline
 $z_{R}$ & \multicolumn{2}{c|}{$1579+264i$} \\
 $(I=1)$ & $g_i$ & $|g_i|$  \\
 \hline
 $\pi\Lambda$ & $ 1.4 + 1.5 i $ & 2.0 \\
 $\pi \Sigma$ & $ -2.2 - 1.5 i $ & 2.7 \\
 ${\bar K} N$ & $ -1.1 - 1.1 i $ & 1.6 \\
 $\eta\Sigma$ & $ 1.2 + 1.4 i $ & 1.9 \\
 $K\Xi$ & $ -2.5 - 2.4 i $ & 3.5 \\
 \hline
\end{tabular}
\label{tab:jido1}
\end{table}

\begin{table}[ht]
\centering \caption{\small Pole positions and couplings to $I=0$
physical states from the model of Ref.~\cite{oller}}
\vspace{0.5cm}
\begin{tabular}{|c|cc|cc|cc|}
 \hline
 $z_{R}$ & \multicolumn{2}{c|}{$1379+27i$} &
\multicolumn{2}{c|}{$1434+11i$} &
 \multicolumn{2}{c|}{$1692+14i$} \\
 $(I=0)$ & $g_i$ & $|g_i|$ & $g_i$ & $|g_i|$ & $g_i$ & $|g_i|$ \\
 \hline
 $\pi \Sigma$ & $-1.76-0.62i$ & 1.87 & $-0.56-1.02i$ & 1.16 & $-0.08-0.32i$ & 0.33 \\
 ${\bar K} N$ & $0.86+0.70i$ & 1.11 & $-1.74+0.63i$ & 1.85 & $0.32+0.41i$ & 0.52 \\
 $\eta\Lambda$ & $0.19+0.33i$ & 0.38 & $-1.20+0.23i$ & 1.23 & $-0.83-0.19i$ & 0.85 \\
 $K\Xi$ & $-0.52-0.19i$ & 0.55 & $-0.20-0.30i$ & 0.36 & $3.87+0.05i$ & 3.87 \\
  \hline
 \end{tabular}
\label{tab:oller0}
\end{table}

\begin{table}[ht]
\centering \caption{\small Pole positions and couplings to $I=1$
physical states from the model of Ref.~\cite{oller}}
\vspace{0.5cm}
\begin{tabular}{|c|cc|cc|}
  \hline
$z_{R}$ & \multicolumn{2}{c|}{$1401+40i$} &
\multicolumn{2}{c|}{$1488+114i$} \\
 $(I=1)$ & $g_i$ & $|g_i|$ & $g_i$ & $|g_i|$  \\
 \hline
 $\pi\Lambda$ & $0.60+0.47 i$ & 0.76 & $0.98+0.84 i$ & 1.3 \\
 $\pi \Sigma$ & $1.27+0.71 i$ & 1.5 & $-1.32-1.00 i$ & 1.7 \\
 ${\bar K} N$ & $-1.24-0.73 i$ & 1.4 & $-0.89-0.57 i$ & 1.1 \\
 $\eta\Sigma$ & $0.56+0.41 i$ & 0.69 & $0.58+0.29 i$ & 0.65 \\
 $K\Xi$ & $0.12+0.05i$ & 0.13 & $-1.63-0.91i$ & 1.9 \\
  \hline
\end{tabular}
\label{tab:oller1}
\end{table}

We now consider the results obtained from the model of
Ref.~\cite{oller}. Making use of their set~I of parameters, which
correspond to a  baryon mass $M_0=1286$ MeV and a meson decay
constant $f=0.798f_{\pi}=74.1$ MeV, both in the chiral limit,
together with a common subtraction constant $a=-2.23$, the results
obtained for $I=0$ and $I=1$ are
displayed in Tables~\ref{tab:oller0} and \ref{tab:oller1},
respectively.

We can see that there is a qualitative agreement among both
models, especially in the case of $I=0$. We observe that the
second resonance with $I=0$ couples strongly to $\bar{K} N$ channel, while
the first resonance couples more strongly to $\pi \Sigma$. The
results for $I=0$ shown in Tables~\ref{tab:jido0}, \ref{tab:oller0} 
resemble much
those obtained in Ref.~\cite{fink} and Ref.~\cite{nieves} where
two resonances are also found close to 1405~MeV, with the one at
lower energies having a larger width than the second and a
stronger coupling to $\pi\Sigma$, while the resonance at higher
energies being narrower and coupling mostly to $\bar{K}N$.
% and that the width of the singlet is much larger than that of the
% octet.

\section{SU(3) considerations}
\label{sec:4}

While in the discussion following Fig.~\ref{fig:tracepole} we have
identified the states as octet and singlet because of their
origin, it is clear that there is some mixing. In fact we can
make a basis transformation and find the coupling of the
resonances to the SU(3) meson--baryon states. We write the matrix
\begin{equation}
\tilde{T}= U^{\dagger} T U
\end{equation}
where $U$ is the unitary matrix of the SU(3) Clebsch--Gordan
coefficients $\langle i,\gamma \rangle$ with $i$ the indices for
the physical states and $\gamma$ denoting a subset of the 
meson--baryon SU(3) states having hypercharge, isospin and isospin
projection compatible with those of the physical ones. Taking the
resonances and couplings found from the model of
Ref.~\cite{bennhold} (Table \ref{tab:jido0}) we find the results
shown in Table \ref{tab:su30}.

\begin{table}[ht]
\centering \caption{\small Couplings of the $I=0$ bound states to
the meson--baryon SU(3) basis states, obtained with the model of
Ref.~\cite{bennhold}} \vspace{0.5cm}
\begin{tabular}{|c|cc|cc|cc|}
\hline
 $z_{R}$ & \multicolumn{2}{c|}{$1390 + 66i$} &
\multicolumn{2}{c|}{$1426 + 16i$} &
 \multicolumn{2}{c|}{$1680 + 20i$} \\
  & \multicolumn{2}{c|}{(evolved singlet)} &
\multicolumn{2}{c|}{(evolved octet $8_s$)} &
 \multicolumn{2}{c|}{(evolved octet $8_a$)} \\
 & $g_\gamma$ & $|g_\gamma|$ & $g_\gamma$ & $|g_\gamma|$ & $g_\gamma$ & $|g_\gamma|$ \\
 \hline
 1 & $2.3+2.3i$ & 3.3 & $-2.1+1.6i$ & 2.6 & $-1.9+0.42i$ &
 2.0 \\
 $8_s$ & $-1.4-0.14i$ & 1.4 & $-1.1-0.62i$ & 1.3 & $-1.5-0.066i$ &
 1.5 \\
 $8_a$ & $0.53+0.94i$ & 1.1 & $-1.7+0.43i$ & 1.8 & $2.6+0.59i$ &
 2.7 \\
 $27$ & $0.25-0.031i$ & 0.25 & $0.18+0.092i$ & 0.21 & $-0.36+0.28i$ &
 0.4 \\
 \hline
 \end{tabular}
\label{tab:su30}
\end{table}

We observe that the physical singlet couples mostly to the singlet
SU(3) state. This means that this physical state has retained
largely the singlet nature it had in the SU(3) symmetric
situation. The same is true for the physical $I=0$  antisymmetric
octet shown in the last column. However, the couplings of the
physical symmetric octet reveal that, due to its proximity to the
singlet state, it has become mostly a singlet with some admixture
of the symmetric and antisymmetric octets.

Alternatively, one can try to determine the SU(3) pole content
of a physical pole by
decomposing the corresponding physical resonance,
denoted by $| A \rangle$, in terms of the {\it bound states} found
in the SU(3) symmetric situation, denoted by $|\mu^\prime\rangle$,
as
 \be \label{su3dec} | A \rangle =
\sum_\mu C^{(A)}_\mu |\mu^\prime \rangle \ , 
\label{eq:bound}
\ee with \be
\label{norm} \sum_\mu |C^{(A)}_\mu|^2 \simeq 1 \ , \ee where the
expansion has been limited to the SU(3) bound states which, in the
present case, are found in the irreducible SU(3) representations
1, $8_a$ or $8_s$. The small admixture of the $27$ representation,
as seen in Table~\ref{tab:su30}, and a possible small contribution of the
continuum states have been neglected in writing Eqs.~(\ref{eq:bound})
 and (\ref{norm}). We note that the state $| A \rangle$ has well
defined isospin ($I=0$ or $1$), third component of isospin and
hypercharge ($Y=0$), the same as the $|\mu^\prime\rangle$ state in
the sum on the right hand side of Eq.~(\ref{su3dec}). For
simplicity, all these quantum numbers are included in the symbols
$A$ and $\mu$. If we further designate by $|\gamma \,MB\rangle$
the basis state belonging to the $\gamma$ SU(3) irreducible
representation and made up by a meson and a baryon, as for
instance the singlet $|1\, MB \rangle = \frac{1}{\sqrt{8}}| K^-p
+ \bar{K}^0n + \pi^+\Sigma^- + \pi^0\Sigma^0 + \pi^- \Sigma^+ +
\eta\Lambda + K^+\Xi^- + K^0\Xi^0 \rangle $, we can write for the
coupling $|A \rangle \rightarrow |\gamma\, MB\rangle\equiv g(A
\rightarrow \gamma)$
 \be g(A \rightarrow
\gamma)=\sum_\mu C^{(A)}_\mu g(\mu^\prime \rightarrow \gamma )~.
\label{tsing} \ee 
In the SU(3) limit $g(\mu^\prime \rightarrow
\gamma )= \delta_{\mu \gamma} g(\gamma^\prime \rightarrow \gamma
)$, hence, up to first order in the SU(3) breaking
parameter, we have
\be g(A \rightarrow \gamma)= C^{(A)}_\gamma
g(\gamma^\prime \rightarrow \gamma )~, \ee 
where the couplings
$g(A \rightarrow \gamma)$ for the $I=0$ resonances are those given
in Table \ref{tab:su30}. Therefore, once the couplings
$g(\gamma^\prime \rightarrow \gamma )$ are known, the equation
above together with the normalization requirement of
Eq.~(\ref{norm}), is sufficient to determine the coefficients
$C^{(A)}_\gamma$ except for a global phase.

In order to calculate the couplings $g(\gamma^\prime\rightarrow
\gamma)$ we consider the  SU(3) limit corresponding to the
physical meson and baryon masses. In order
to do that we take the Gell-Mann-Okubo mass relations for the
lightest octet of pseudoscalars and baryons for calculating the
common meson and baryon masses in the SU(3) limit, see e.g.
Refs.~\cite{georgi,licht}. These result to be $M_B=1150$ MeV for
the baryons and $m_0=413$ MeV for the mesons.  The pole positions
of the singlet and octet ($8_s$ and $8_a$), together with
the corresponding residua,
obtained from the two models studied in this
work are
\begin{equation}
\begin{array}{|l|c|c|}
\hline
  & \mbox{Ref.~\cite{bennhold}} & \mbox{Ref.~\cite{oller}} \\
\hline
\mbox{singlet pole} & (1486,0) & (1447,0) \\
\mbox{octet pole} & (1556,0)  & (1516,0) \\
g(1^\prime \rightarrow 1) & (-3.29,0) & (-3.05,0) \\ 
g(8^\prime \rightarrow 8) & (-1.87,0) & (-2.65,0)  
 \\
\hline
 \end{array} \ .
\label{tab:su3}
\end{equation}

Taking the $g(A \rightarrow \gamma)$ couplings
from Table~\ref{tab:su30}, one obtains the following coefficients

 \begin{equation}
   \begin{array}{|c|l|l|l|l|l|l|}
   \hline
\hbox{Pole} & C_1 & C_{8_a}/C_1 & C_{8_s}/C_{1} &
|C_1|^2 & |C_{8_a}|^2 & |C_{8_s}|^2 \\
\hline
1390 + 66i & 0.73 & (  0.41,  0.12)& ( -0.43,  0.34) & 0.53& 0.18 & 0.29\\
\hline
1426+16i & 0.56 & (0.62,  0.27)  & (0.20,  0.45)& 0.31 & 0.45& 0.24\\
\hline
1680+20i & 0.34 & (-0.75, -0.35) &  ( 0.43,  0.11) & 0.12  & 0.68& 0.20\\
\hline
\end{array} \ ,
\label{tabi0jido}
\end{equation}   
which again confirms that the first pole is mostly a singlet, the third one 
mostly an antisymmetric octet and the second one has become an even mixture of
the three SU(3) poles.
Following the same steps but now for the $I=1$ pole found in the model of 
Ref.~\cite{bennhold}, one obtains the coefficients
\begin{equation}
   \begin{array}{|c|l|l|l|l|}
   \hline
   \hbox{Pole} & C_{8_a} & C_{8_s}/C_{8_a} & |C_{8_a}|^2 &
|C_{8_s}|^2 \\
\hline
1579+264i & 0.55 &  (0.83,  0.14) &  0.30 & 0.70\\
\hline
\end{array} \ ,
\label{tabi1jido}
\end{equation}
which show that this pole has retained the $8_s$ nature it had in the SU(3) symmetric
situation (see Eq.~(\ref{eq:deviations}) in appendix B).
Similarly, the results obtained from the model of Ref.~\cite{oller} are
\be
\begin{array}{|c|l|l|l|l|l|l|}
\hline
\hbox{Pole} & C_1 & C_{8_a}/C_1 & C_{8_s}/C_{1} &
|C_1|^2 & |C_{8_a}|^2 & |C_{8_s}|^2 \\
\hline
1379+27i & 0.96 & (0.15,0.11) & (0.15,-0.19) & 0.92 & 0.03 & 0.05\\
\hline
1434+11i & 0.49 & (0.64,0.77) & (0.71,1.28) & 0.24 & 0.24 & 0.52\\
\hline
1692+14i & 0.48 &  (1.58,0.37) & (0.78,0.16) & 0.23 & 0.63 & 0.14\\
\hline
\end{array}
\label{tabi0oller} \ , \ee
for the $I=0$ poles and
\be
\begin{array}{|c|l|l|l|l|}
\hline
\hbox{Pole} & C_{8_a} & C_{8_s}/C_{8_a} & |C_{8_a}|^2 &
|C_{8_s}|^2 \\
\hline
1401+40i & 0.81 &  (0.72,0.07) &  0.66 & 0.34\\
\hline
1488+114i & 0.59 & (1.37,-0.06) & 0.35 & 0.65\\
\hline
\end{array}
\label{tabi1oller} \ ,\ee 
for the $I=1$ ones. In this model, for which SU(3) breaking is less pronounced,
the physical poles retain better the irreducible SU(3) nature they had in the
SU(3) symmetric situation.

If the basis transformation of Eq.~(\ref{eq:bound}) was complete,
the sum of the moduli square of
the coefficients should also be one along a column, which
corresponds to expanding the SU(3) eigenstates in the physical 
basis of resonances.
Within 20\% this is true for all the
columns in the tables of Eqs.~(\ref{tabi0jido}), 
(\ref{tabi0oller}) and (\ref{tabi1oller}),
except for the one corresponding to the SU(3) singlet state in the
model of Ref.~\cite{oller}, which
is overestimated in a $40\%$ [see Eq.~(\ref{tabi0oller})].

\section{Influence of the poles on the physical observables}
\label{sec:5}

\begin{figure}
\epsfxsize=12cm \centerline{\epsfbox{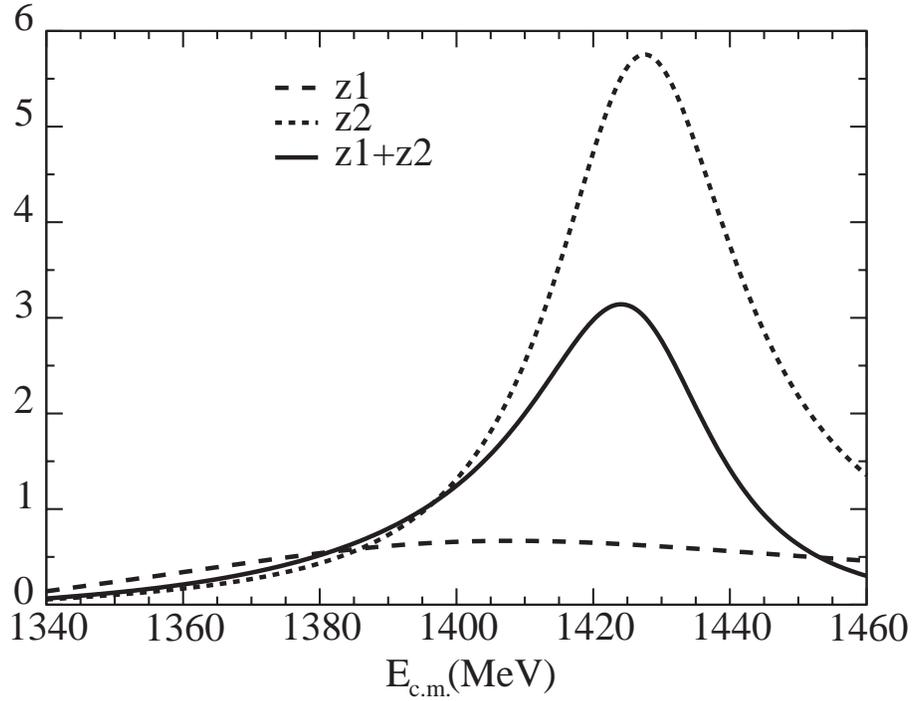}} 
\caption{$z_{1}$: the modulus
square of the first term in Eq.~(\ref{eq:kbarNamp}) multiplied by
$p_{\pi}$. $z_{2}$: Same for the second term in
Eq.~(\ref{eq:kbarNamp}). $z_{1}+z_{2}$: Same for the coherent sum
of the two terms. \label{fig:dist1}}
\end{figure}

\begin{figure}
\epsfxsize=12cm \centerline{\epsfbox{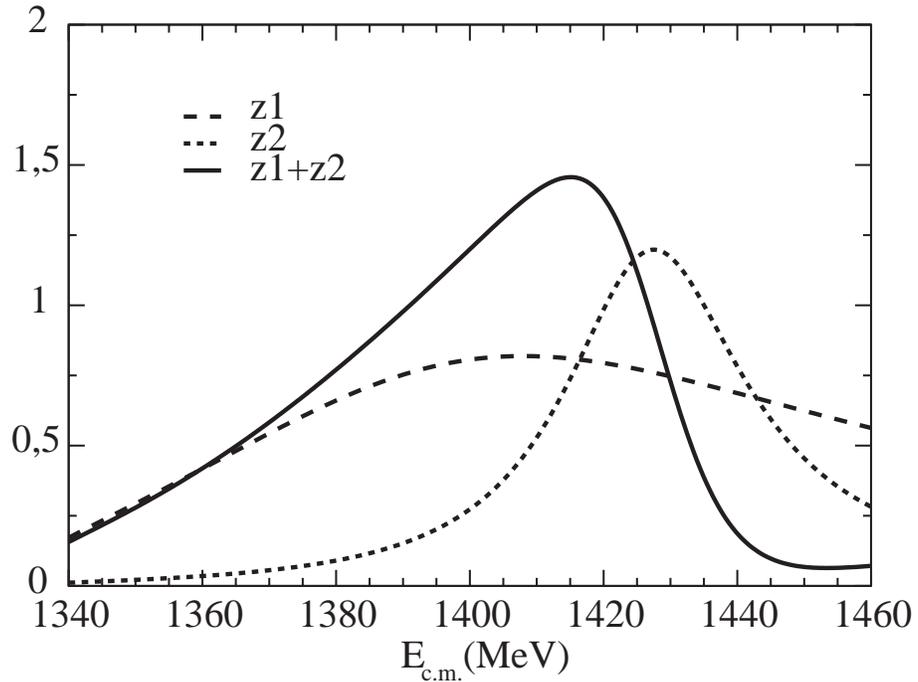}} 
\caption{Same as
Fig.~\ref{fig:dist1} but for the terms in Eq.~(\ref{eq:pisigamp}).
\label{fig:dist2}}
\end{figure}

It is  important to see what would happen in an actual experiment.
In order  to see this, we first make a qualitative and intuitive
exercise and then compare to the numerical results obtained using
the model of Ref.~\cite{bennhold}. Consider now two resonances,
called $R_{1}$ and $R_{2}$, corresponding to the physical singlet
and symmetric octet states. Take the complex couplings of Table~\ref{tab:jido0} 
and construct the following amplitudes
\begin{eqnarray}
&& g_{\bar{K}N}^{R_{1}} {1 \over W - M_{R_{1}} + i\Gamma_{R_{1}}/2}
          g_{\pi\Sigma}^{R_{1}} +
      g_{\bar{K}N}^{R_{2}} {1 \over W - M_{R_{2}} + i\Gamma_{R_{2}}/2}
          g_{\pi\Sigma}^{R_{2}}\ ,
          \label{eq:kbarNamp}\\
&& g_{\pi\Sigma}^{R_{1}} {1 \over W - M_{R_{1}} + i\Gamma_{R_{1}}/2}
        g_{\pi\Sigma}^{R_{1}} +
      g_{\pi\Sigma}^{R_{2}}  {1 \over W - M_{R_{2}} + i\Gamma_{R_{2}}/2}
        g_{\pi\Sigma}^{R_{2}} \,\, .
        \label{eq:pisigamp}
\end{eqnarray}
This would be basically equivalent to  the amplitudes $T_{\bar{K}N
\rightarrow \pi\Sigma}$ and $T_{\pi\Sigma
  \rightarrow \pi \Sigma}$, respectively.
In Figs.~\ref{fig:dist1} and \ref{fig:dist2} we plot the modulus
square of these two quantities multiplied by the $\pi\Sigma$
momentum as a function of the energy. We also show the
contribution of each resonance by itself (dotted and dashed
lines). In both cases one only sees one resonant shape (solid
line) but the simulated $T_{\pi\Sigma \rightarrow \pi\Sigma}$
amplitude in Fig.~\ref{fig:dist2} produces a resonance at a lower
energy and with a larger width. This case reproduces very well the
nominal experimental $\Lambda(1405)$, both in the position and
width. Actually, what is done in Ref.~\cite{angels} to get the
shape of the $\Lambda(1405)$ is precisely to plot the invariant
mass distribution of the $\pi \Sigma$ states according to the
expression
\begin{equation}
 {d \sigma \over dM_{i}} = C |T_{\pi \Sigma \rightarrow \pi \Sigma}|^{2}
     q_{\rm c.m.} \ ,
\label{eq:ps}
\end{equation}
where $C$ is a constant.
However, if the invariant mass distribution of the $\pi\Sigma$
states were dominated by the $\bar{K}N \rightarrow \pi\Sigma$
amplitude, then the second resonance $R_{2}$ would be weighted
more, since it has a stronger coupling to the $\bar{K}N$ state,
resulting into an apparent narrower resonance peaking at higher
energies as illustrated in Fig.~\ref{fig:dist1}. In order to avoid
such ambiguities, in the calculation of the $\pi\Sigma$ invariant
mass distribution of Ref.~\cite{oller} a coupled channel scheme
was implemented from the onset, not only to obtain the strong
S-wave $T$-matrix, but also in the production mechanism of the
$I=0$ $\pi\Sigma$ state, involving initial $\pi\Sigma$ and
$\bar{K}N$ states. The weight of the $K^- p$ channel relative to
the $\pi^-\Sigma^+$ one in the production mechanism, obtained in
Ref.~\cite{oller} from a fit to the data, 
is $r_{K^- p}/r_{\pi^-\Sigma^+}=1.416$, being
the same for any other $\bar{K}N$ or $\pi\Sigma$ channel, since
the source has isospin $I=0$. The usual Eq. (\ref{eq:ps}) is
obtained from this formalism when $r_{\bar{K}N}=0$. However, in
Ref.~\cite{oller} a good description of the data is achieved when
the $\bar{K}N$ channel is allowed to participate directly in the
production mechanism as well ($r_{\bar{K}N}\neq 0$) so that
finally only {\it one effective} resonant shape is visible,
despite of being determined by two rather narrow poles 
(see Table~\ref{tab:oller0}). Let us note that in
Refs.~\cite{angels,bennhold} the first pole in Table
\ref{tab:jido0} has a width twice as large as that of the
corresponding pole of Ref.~\cite{oller}, giving rise to a good
reproduction of the $\pi\Sigma$ invariant mass distribution in
terms of only the $T_{\pi\Sigma \rightarrow \pi\Sigma}$ amplitude,
the one appearing in Eq.~(\ref{eq:ps}). This is no longer possible
in Ref.~\cite{oller} because the first pole is much narrower and
the aforementioned coupled channel formalism is unavoidable in
order to find agreement with data. For further details related to
the formalism see Refs.~\cite{oller,meiss}. This discussion 
makes clear that a theoretical investigation
of the rates $r_{\bar{K}N}$ and $r_{\pi\Sigma}$ from 
different reactions is a key ingredient for a complete
understanding of the dynamics of the $\Lambda(1405)$ state.

\begin{figure}
\epsfxsize=12cm
\centerline{\epsfbox{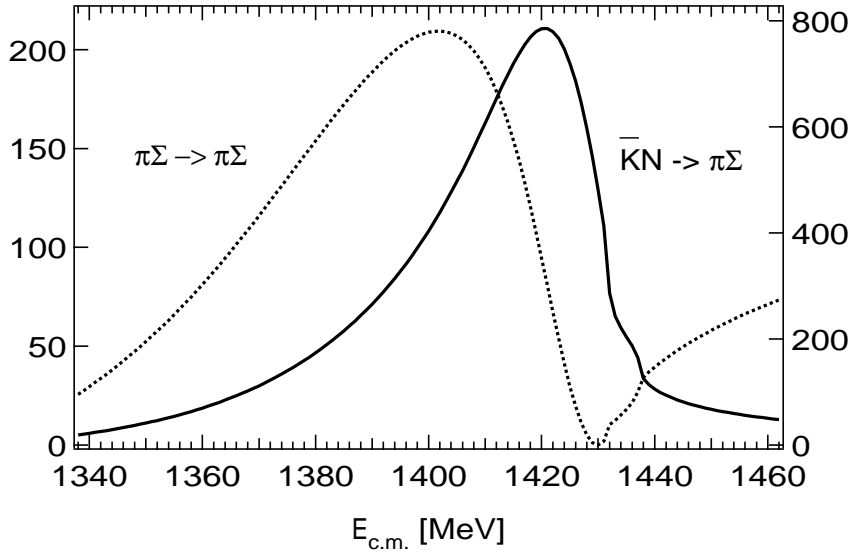}}
\caption{The $\pi\Sigma$ mass distributions with $I=0$ constructed from
  the $\bar{K}N\rightarrow \pi\Sigma$ and $\pi\Sigma \rightarrow \pi\Sigma$
  amplitudes. The solid and dashed lines denote
  $|T_{\bar{K}N\rightarrow \pi\Sigma}|^{2} q_{\pi}$ and
  $|T_{\pi\Sigma\rightarrow \pi\Sigma}|^{2} q_{\pi}$, respectively.
  Units are arbitrary.
\label{fig:massdist}}
\end{figure}

\begin{figure}
\epsfxsize=12cm \centerline{\epsfbox{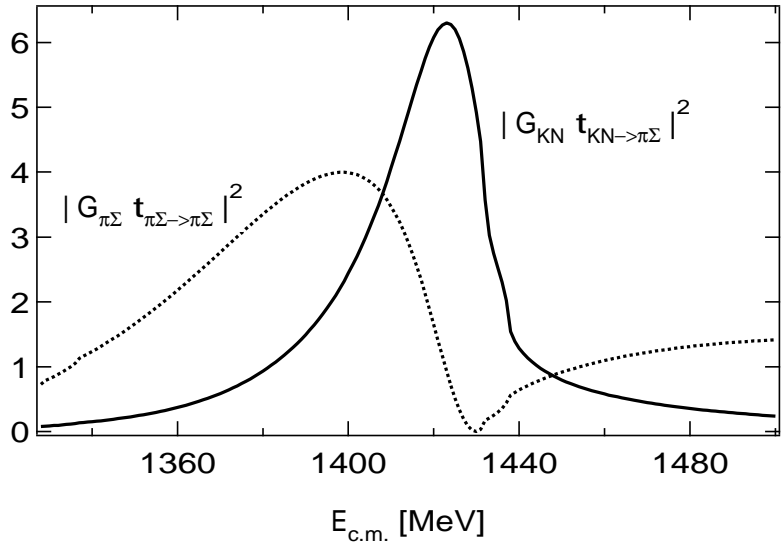}} 
\caption{Modulus squared of the
loop function $G_l$ times the amplitudes, simulating a reaction with
the resonance initiated by the $\pi\Sigma$ (dashed line) or the
$\bar{K}N$ (solid lines) channels. \label{fig:prob}}
\end{figure}

We now turn to more realistic calculations in which we use the
amplitudes generated in the chiral unitary approach of
Ref.~\cite{bennhold}, rather than the approximated ones of
Eqs.~(\ref{eq:kbarNamp}) and (\ref{eq:pisigamp}). In
Fig.~\ref{fig:massdist}, we show the $\pi\Sigma$ invariant mass
distributions constructed from the $\pi\Sigma \rightarrow \pi
\Sigma $ (dotted line) and $\bar{K}N \rightarrow \pi\Sigma$ (solid
line) amplitudes.
% using the amplitudes generated in the approach of Ref.~\cite{bennhold} rather than
As we expected, in the $\bar{K}N \rightarrow \pi \Sigma$ case, the
resonance peaks at a higher energy and shows a narrower width. In
Fig.~\ref{fig:prob} we include the corresponding loop function
$G_l$ and plot the quantities $|G_{\pi\Sigma} T_{\pi\Sigma
\rightarrow \pi\Sigma}|^{2}$ (dotted line) and $|G_{\bar{K}N}
T_{\bar{K}N\rightarrow \pi\Sigma}|^{2}$ (solid line), which would
appear in production processes of the $\Lambda(1405)$ assuming the
build--up of the resonance in the multiple scattering is
initiated by a $\pi\Sigma$ or $\bar{K}N$ state, respectively. It
is interesting to see that in the case of the $\bar{K}N$ initial
state the peak is narrower and appears at higher energy. This is another
important result of this paper.

%%%%%%

We can see that the results of the full calculation are very
similar to those obtained with the qualitative model which allowed
us to see that, in spite of starting from two quite different
resonance poles, the result in an experiment would be to see a single
resonant form. Yet, the shape and width could be different in
different processes, depending on the weight that is given to the
physical channels that initiate the resonance.

It is clear that,  should there be a reaction which forces the
initial channels to be $\bar{K}N$, then this would give more
weight to the second resonance, $R_{2}$, and hence produce a
distribution with a shape corresponding to an effective resonance
narrower than the nominal one and at higher energy. Such a case
indeed occurs in the reaction $K^- p \to \Lambda(1405) \gamma$
studied theoretically in Ref.~\cite{nacher}.  It was shown there
that since the $K^- p$ system has a larger energy than the
resonance, one has to lose energy emitting a photon prior to the
creation of the resonance and this is effectively done by the
Bremsstrahlung from the original $K^-$ or the proton.  Hence the
resonance is initiated from the  $K^- p$ channel.

\section{Conclusions}
\label{sec:6}

 In this paper we have investigated the poles appearing in the
 meson--baryon scattering matrix for strangeness $S=-1$ within a 
coupled--channel chiral
unitary
 approach, using two different methods for breaking the SU(3)
symmetry which have been used in the literature.

   In both approaches some resonances are generated dynamically from the
 interaction of the octet of pseudoscalar mesons with the octet of the
$1/2^+$
 baryons.  The underlying SU(3) structure of the Lagrangians implies
that, from
 the combination of the two original octets, a singlet and two octets of
 dynamically generated resonances should appear, but the dynamics of the
problem makes the two octets degenerate in the case of
exact SU(3) symmetry.  
The same chiral Lagrangians have mechanisms for  chiral
 symmetry breaking which have as a consequence that the degeneracy is broken
 and two distinct octets appear. In
Ref.~\cite{oller} the interaction kernel $V$ is calculated
strictly at lowest order in the chiral expansion (with a common
baryon mass $M_0$), including consistently at this order the direct and crossed one baryon 
exchange contributions, while Refs.\cite{angels,bennhold} neglect the latter and 
utilize the physical baryon masses. In addition in ref.\cite{oller} a unique substraction 
constant is employed, as corresponds to the SU(3) limit, see appendix A, while in 
Refs.\cite{angels,bennhold} the subtraction constants are different from channel to channel 
and parameterized in terms of a cut-off. As a consequence the SU(3) 
breaking as well as the interacting kernel $V$ itself are different between 
Refs.\cite{angels,bennhold} and Ref.\cite{oller}.  Although the SU(3) symmetry breaking 
mechanisms can be more general, and can be systematically included by the evaluation 
of $V$ to increasingly higher orders in a chiral expansion as shown in 
ref.\cite{oller}, the two
approaches followed in the present study, despite their differences,
 have shown very good agreement with the experimental observables
in different $\bar{K}N$ reactions, thus giving  support to our conclusions. 
 
  The breaking of the octet degeneracy has a a consequence that, in the
physical limit, one of the
  $I=0$ octet poles appears
 quite close to the singlet pole,
and both of them are
 very close to the nominal $\Lambda(1405)$.  These two resonances are quite
 close but different, the one at lower energies with a larger width and
a stronger 
 coupling to the $\pi \Sigma$ states than the one at higher energies, which
 couples mostly to the $\bar{K}N$ states.  This is the main finding of the
 present work, thus we conclude that there is not just one single
$\Lambda(1405)$
 resonance, but {\em two}, and that what one sees in experiments is a
{\em superposition} of these two states.   
 
  Another interesting finding of the paper is the suggestion that it is
 possible to find out the existence of the two resonances by performing
 different experiments, since in different experiments the weights by which the two
 resonances are excited are different.  In this respect we call the
attention to one reaction,  $K^-p \to \Lambda(1405) \gamma$, which gives much
 weight to the resonance which couples strongly to the $\bar{K}N$ states
and, hence, leads to a peak structure in the invariant mass distributions 
 which is
 narrower and appears at higher energies than the experimental
$\Lambda(1405)$
 peaks observed in hadronic experiments performed so far.
 
   The two different approaches discussed in the paper to break SU(3)
symmetry have also served
to give an idea about  the uncertainties of
our predictions, resulting from higher order chiral terms in the
calculation of interaction kernel $V$ not considered here.  The relatively good 
agreement of the two approaches
 gives us confidence about the conclusions of the paper concerning the
existence
 of the two  $\Lambda(1405)$ resonances and their different coupling to the
 meson--baryon states. This important theoretical finding should
stimulate new
 experiments exciting the $\Lambda(1405)$ resonance, as well as new
 analyses to unravel the double pole structure of the peaks seen in the
 reactions.
                         
\section*{Acknowledgments}
%We are very grateful to ... for
%useful discussions.  
One of us, E.O.,  would like to acknowledge useful discussions  with B. Holstein and 
J. Gasser on issues of this paper.  We would also like to acknowledge the encouragement 
of H. Toki for us to make this work and his collaboration in former works which 
stimulated us to do the present one. This work is partially supported by DGICYT
projects BFM2000-1326, BFM2001-01868, FPA2002-03265,
the EU network EURIDICE contract
HPRN-CT-2002-00311,
and the Generalitat de Catalunya project
2001SGR00064.
D.J. would like to acknowledge the support of Japanese
Ministry of Education, Culture, Sports, Science and Technology
to stay at IFIC, University of Valencia, where part of this work was
done.

\appendix{}
\section{Subtraction constants in the SU(3) limit}
\label{app:A}
\def\theequation{\Alph{section}.\arabic{equation}}
\setcounter{equation}{0}

In this appendix we want to show how the $a_i$ subtraction
constants, present in the $G_i(s)$ functions, become equal in the
SU(3) limit, which is assumed to hold in all this appendix.

 Each SU(3) irreducible representation
decouples and then, analogously to Eq.~(\ref{eq:bs1}) which is
given for the physical channels, we will have for each irreducible
representation $\gamma$ the amplitude: \be \label{tsu3}
T_\gamma=\frac{V_\gamma}{1-V_\gamma G_\gamma}~, \label{tnsu3} \ee
where now all the functions in the previous formula are just
numbers due to the decoupling of different irreducible
representations. The subtraction constant of $G_\gamma$ is denoted
by $a_\gamma$.

Nevertheless, we can still use Eq.~(\ref{eq:bs1}) and deduce the
$T_\gamma$ amplitudes by an orthonormal transformation.
Correspondingly, we are taking the SU(3) Clebsch-Gordan
coefficients real so that they satisfy: \ba
\sum_i \langle i,\gamma\rangle \langle i,\mu\rangle&=&\delta_{\gamma \mu}~,\nn\\
\sum_\gamma \langle i,\gamma \rangle \langle
j,\gamma\rangle&=&\delta_{ij}~, \label{nor2} \ea where the latin
indexes refer to the physical channels and the greek ones to the
SU(3) eigenstates. Eq.~(\ref{eq:bs1}), which applies to the
physical channels, can be rewritten as: \be
T_{ij}=V_{ij}+\sum_{k,k'} V_{ik}G_{kk'} T_{k'j}~, \ee where
$G_{kk'}$ is a diagonal matrix, $G_{kk'}=G_k \delta_{kk'}$, as
discussed in section \ref{sec:2}. Hence, by making a change of
basis, we find: \be \sum_{ij}\langle i,\gamma\rangle A_{ij}\langle
j,\mu \rangle= A_{\gamma \mu} = A_\gamma \delta_{\gamma\mu}~, \ee
with $A$ standing for $V$, $G$ or $T$, and we have used the fact
that in the SU(3) limit $V$, $G$ and $T$ are singlet
operators.
Inverting the previous equation for $A\equiv G$ we find:
\be G_{kk'}=\sum_\mu \langle k,\mu\rangle G_\mu \langle
k',\mu\rangle~, \ee and thus, \ba \sum_{k'}G_{kk'}\langle
k',\gamma\rangle& =&\sum_{k'}\sum_\mu \langle k,\mu \rangle G_\mu
\langle k',\mu \rangle \langle k',\gamma \rangle
\nn\\
&=&\langle k,\gamma \rangle G_\gamma~. \ea Using now that
$G_{kk'}$ is diagonal, one has,
 \be G_k \langle k,\gamma\rangle
=G_\gamma \langle k,\gamma\rangle~. \ee
Since this relation is
valid for all $k$ and all $\gamma$, it is sufficient to take
several physical states which have components in different SU(3)
representations\footnote{Let us note that since $G$ is a SU(3)
singlet
  operator, its matrix elements in the SU(3) basis are $G_\gamma$ times
the identity matrix for
 the states belonging to the same irreducible representation.},
to see that all the $G_\gamma$ functions must be
the same, and as a consequence, all the $G_k$ are also  equal.
Equivalently,
the subtraction constants $a_\gamma$ turn out to be the
same in the SU(3) limit and, consequently, the
subtraction constants $a_k$ are independent of the physical channel.

\section{Deviations of the pole positions from the SU(3) limit} 
\label{app:B}
\def\theequation{\Alph{section}.\arabic{equation}}
\setcounter{equation}{0}

In this appendix we would like to show the direction of deviation
of the pole positions from the SU(3) symmetric limit when an
infinitesimal SU(3)
breaking is assumed.
In the SU(3) symmetric limit, we find two bound states, one corresponds
to the singlet and the other corresponds to the degenerate octets
($8_{s}$ and $8_{a}$).
Due to the SU(3) breaking effects, 
the isospin $I=0,1$ states of the two octets ($8_s$ and
$8_a$) split apart in four states as shown 
in Fig.~\ref{fig:tracepole}.
 
The energies of the bound states are calculated as the solutions $z$
of the secular equation:
\begin{equation}
\det \left[ 1-V(z) G(z) \right] = 0 \ . \label{eq:det}
\end{equation}
In the SU(3) symmetric limit, the scattering amplitude is expressed as
a diagonal matrix
in the basis of the SU(3) irreducible representations, as given in
Eq.~(\ref{tsu3}).
For the diagonal matrix the determinant is written in the product from:
\begin{equation}
   \prod_{\gamma} \left( 1-V_{\gamma}(z_{0}^{\gamma})
    G_{0}(z_{0}^{\gamma}) \right) = 0 \ ,
\end{equation}
where $z_{0}^{\gamma}$ denotes the solution for the irreducible
representation $\gamma$
in the SU(3) symmetric limit and the loop integral $G_{0}$ is
independent of the
irreducible representation in this limit, as discussed in
appendix
\ref{app:A}.   

Now let us introduce an infinitesimal breaking of the SU(3) symmetry to
the $G$ function through the subtraction constants and the masses of
baryons
and mesons as in Eq.~(10). We expand the $G$ function
in term of the SU(3) breaking parameter $x$:
\begin{equation}
G_{\gamma}  = G_{0} +    \tilde G_{\gamma}\delta x \ ,
\end{equation}
where $\delta x \ll 1$ and
\begin{equation}
   \tilde G_{\gamma} = \left. {\partial G_{\gamma} \over \partial
x}\right|_{x=0}
  = \sum_{i}\langle i,\gamma \rangle
   \left. {\partial G_{i} \over \partial x}\right|_{x=0} \langle i,
\gamma \rangle
   \ .
\end{equation}
In general, the $G$ function in the SU(3) basis has off-diagonal
components when the
SU(3) breaking effects are introduced. Nevertheless, for small $\delta
x$, such
off-diagonal components contribute to higher orders when the energies
of the bound
states are calculated using Eq.~(\ref{eq:det}). Therefore, the equation
determining the positions of
the poles is obtained again in the decoupled form:
\begin{equation}
   \prod_{\gamma} \left( 1-V_{\gamma}(z^{\gamma})
    \{G_{0}(z^{\gamma}) + \tilde G(z^{\gamma})\delta x \} \right) = 0 \ .
\end{equation}
 
Regarding $\tilde G \delta x $ as a perturbation, we can calculate
deviations
of the positions of the poles due to a small SU(3) breaking effect.
Writing $z^{\gamma}= z_{0}^{\gamma}+ \epsilon_{\gamma}$,  we
obtain
\begin{equation}
  \epsilon_{\gamma} = \left. -{V_{\gamma} \tilde G_{\gamma} \over
  V^{\prime}_{\gamma} G_{0} + G^{\prime}_{0}V_{\gamma}} \delta x
  \right|_{z=z_{0}^{\gamma}} \ ,
\end{equation}
where $V^{\prime}$ and $G_{0}^{\prime}$ are derivatives of the $V$ and
$G$ functions
with respect to the energy $z$. In the case of the octets, the sign of
$\epsilon_{\gamma}$
depends only on $\tilde G_{\gamma}$. Here we show the numerical
results of the deviations at $\delta x = 0.05$:   
\begin{equation}
   \begin{array}{lll}
     \epsilon_{1} = 1.2 ~{\rm [MeV]} ~~~ &
     \epsilon_{8_{s}, I=0} =  -0.30 ~{\rm [MeV]}~~~&
     \epsilon_{8_{a}, I=0} =  5.7 ~{\rm [MeV]}\\
    ~~~&\epsilon_{8_{s}, I=1} = 4.1 ~{\rm [MeV]}~~~&
     \epsilon_{8_{a}, I=1} = -1.6 ~{\rm [MeV]}
   \end{array} \ ,
\label{eq:deviations}
\end{equation} 
which explain the deviations in the positions of the poles with increasing
$x$ observed in Fig.~\ref{fig:tracepole}.

\end{document}